\documentclass{article}
\usepackage{amsmath,amsfonts,bm}

\def\eqref#1{equation~\ref{#1}}
\def\1{\bm{1}}

\def\rc{{\textnormal{c}}}

\def\ry{{\textnormal{y}}}
\def\rz{{\textnormal{z}}}

\def\rvx{{\mathbf{x}}}
\def\rvy{{\mathbf{y}}}
\def\rvz{{\mathbf{z}}}

\def\vy{{\bm{y}}}
\def\vz{{\bm{z}}}

\DeclareMathAlphabet{\mathsfit}{\encodingdefault}{\sfdefault}{m}{sl}
\SetMathAlphabet{\mathsfit}{bold}{\encodingdefault}{\sfdefault}{bx}{n}

\usepackage{spconf,amsmath,graphicx}
\usepackage{environ}         % provides \BODY
\usepackage{balance}
\usepackage{etoolbox}        % provides \ifdimcomp
\usepackage{cite}
\usepackage[lofdepth,lotdepth]{subfig}

\title{Deep Clustering of Compressed Variational Embeddings}
\name{Suya Wu$^{1}$ \qquad Enmao Diao$^{1}$ 
\qquad Jie Ding$^{2}$ \qquad Vahid Tarokh$^{1}$\thanks{This work was supported in part by Office of Naval Research Grant No. N00014-18-1-2244.}}

\address{$^{1}$ Department of Electrical and Computer Engineering, Duke University, Durham, NC 27701, USA\\
$^{2}$ School of Statistics, University of Minnesota-Twin Cities, Minneapolis, MN 55455, USA
}

\newlength{\myl}
\let\origequation=\equation
\let\origendequation=\endequation

\RenewEnviron{equation}{
  \settowidth{\myl}{$\BODY$}                       % calculate width and save as \myl
  \origequation
  \ifdimcomp{\the\linewidth}{>}{\the\myl}
  {\ensuremath{\BODY}}                             % True
  {\resizebox{\linewidth}{!}{\ensuremath{\BODY}}}  % False
  \origendequation
}
\begin{document}
%\ninept
%
\maketitle
\begin{abstract}
Motivated by the ever-increasing demands for limited communication bandwidth and low-power consumption, we propose a new methodology, named joint Variational Autoencoders with Bernoulli mixture models (VAB), for performing clustering in the compressed data domain. The idea is to reduce the data dimension by Variational Autoencoders (VAEs) and group data representations by Bernoulli mixture models (BMMs). Once jointly trained for compression and clustering, the model can be decomposed into two parts: a data vendor that encodes the raw data into compressed data, and a data consumer that classifies the received (compressed) data. In this way, the data vendor benefits from data security and communication bandwidth, while the data consumer benefits from low computational complexity. To enable training using the gradient descent algorithm, we propose to use the Gumbel-Softmax distribution to resolve the infeasibility of the back-propagation algorithm when assessing categorical samples.
\end{abstract}
\section{Introduction}
\label{sec:intro}
Clustering is a fundamental task with applications in medical imaging, social network analysis, bioinformatics, computer graphics, etc. Applying classical clustering methods directly to high dimensional data may be computational inefficient and suffer from instability. Many recent papers have shown that clusters for high dimensional data lie only in subsets of the full space and good data representations are beneficial to clustering \cite{Xie, AE-Data-Song,Yang,Li,Bo}. Deep Embedded Clustering (DEC) was proposed to jointly learn feature representations and assign clusters using a class of feedforward artificial neural networks \cite{Xie}. It achieved impressive performances on clustering tasks and it is often treated as the baseline for deep clustering methods. With the same motivation, the authors of \cite{Yang} learned Joint Unsupervised LEarning (JULE) to combine the agglomerative clustering with a Convolutional Neural Network (CNN) and formulated them as a recurrent process. In \cite{Bo}, an identification criterion was proposed to address identifiability issues for nonlinear mixture models. Variational Deep Embedding (VaDE) applied a mixture of Gaussian models as the prior distribution of latent representations in Variational Autoencoders (VAEs), therefore modeling the data generative procedure jointly with the clusters' distribution \cite{Vade}. 

On the other hand, lossy compression achieves a high compression ratio to reduce computation and communication costs. In recent works, VAEs was applied for lossy image compression and achieved comparable results \cite{Zhou,Johannes,Lucas}. The authors of aforementioned papers presented an end-to-end image compression framework for low bit-rate compression by applying VAEs with a Laplacian distribution \cite{Zhou}. A similar method was described for effective capturing of spatial dependencies in the latent representations based on VAEs in \cite{Johannes}. It has been shown that VAEs have the potential to address an increasing need for flexible image compression \cite{Lucas}. Taken together, prior research provides evidence that a better model fitting leads to better compression performance, and consequently enables a more accurate clustering assignment. 

Performing clustering on compressed data is a potential solution to problems arising in storage, computing, and communicating unstructured and unlabelled image collections. In contrast with previous works, where the proposed methods learned the data representation either specifically for clustering or for compression, we explore both tasks simultaneously by a new method, namely joint Variational Autoencoders and Bernoulli mixture models (VAB). This performs deep clustering on binary representations of data with state-of-the-art performance at a high-compression regime. The model is trained in two steps: First, Variational Autoencoders (VAEs) are jointly trained with Bernoulli mixture models (BMMs), where a mixture of Bernoulli models provides a probabilistic distribution of latent representations. Subsequently, the classifier is updated by Bernoulli distributed samples produced in the first step. This is optimized by the loss function consists of a reconstruction loss and a clustering loss. Learning discrete representations with neural network architectures is challenging because of the inability to backpropagate through non-differentiable samples. In our work, we propose to use the Gumbel-Softmax \cite{cVAE}, which provides a differentiable sampling mechanism that trains the neural network with a categorical reparameterization trick. This framework explores the connection between directed probabilistic models and compressed data representations, therefore making it possible to consider interpretable and computationally efficient binary code. To the best of our knowledge, what we present is the first methodology for simultaneous data compression and clustering in compressed domains. 
\section{Methods}
\label{sec:method}
\subsection{The generative model}
Considering the dataset $\rvx$ with $N$ identically independently distributed (i.i.d) samples $\{x_i\}_{i=1}^N$ and $x_i\in \mathbf{R}^d$, we assume that the data is generated by some random process, involving an unobserved Bernoulli random variable $\rvz$ which belongs to one of $k$ classes. The joint distribution is formulated as
\begin{equation}
    p_{\boldsymbol{\theta}}(\rvx,\rvz,\rc)=p_{\boldsymbol{\theta}}(\rc)p_{\boldsymbol{\theta}}(\rvz|\rc)p_{\boldsymbol{\theta}}(\rvx|\rvz,\rc),
\end{equation}
where $\boldsymbol{\theta}$ stands for the generative model parameters. It says that an observe $\rvx$ is generated from a latent variable $\rvz$, and $\rvz$ follows the mixture distributions with respect to ($w.r.t.$) the classes variable $\rc$.  Their distributions are described as:
\begin{align}
    \rvx \sim Bernoulli(\boldsymbol{\mu_{x}}) \;or\;\rvx \sim N(\boldsymbol{\mu_{x}},\boldsymbol{\sigma_{x}}\mathbf{I}) \\
\rvz \sim Bernoulli(\boldsymbol{\mu_{z}})\\
% \;or\;\rvz \sim N(\boldsymbol{\mu_z},\boldsymbol{\sigma_z}\mathbf{I}) \\
\rc \sim Categorical(\boldsymbol{\pi}).
\end{align}
Along with this generative process, we assume \begin{equation}
    p_{\boldsymbol{\theta}}(\rvx|\rvz,\rc)=p_{\boldsymbol{\theta}}(\rvx|\rvz),
\end{equation}
which means that $\rvx$ and $\rc$ are independent conditioning on $\rvz$. We define a recognition model $q_{\boldsymbol{\phi}}(\rvz,\rc|\rvx)$ as the variational approximation under the KL-divergence to the intractable posterior $p_{\boldsymbol{\theta}}(\rvz,\rc|\rvx)$ and $\boldsymbol{\phi}$ stands for the recognition model parameters.
\subsection{The variational lower bound}
 The log-likelihood of $N$ observed data $\rvx$ is \begin{align*}
     \log p(\rvx^{(1)},\rvx^{(2)},\cdots,\rvx^{(N)})=\sum_{i=1}^N\log p(\rvx^{(i)}).
 \end{align*} Each element of the whole loglikelihood for the observed data is written as
\begin{align*}
    \log p(\rvx^{(i)})=D_{\textsc{KL}}(q_{\phi}(\rvz,\rc|\rvx^{(i)})||p_{\boldsymbol{\theta}}(\rvz,\rc|\rvx^{(i)})) +  \mathcal{L}(\boldsymbol{\theta},\boldsymbol{\phi};\rvx^{(i)}),
\end{align*}
where the first term is the KL-divergence of the learned posterior distribution $q_{\phi}(\rvz,\rc|\rvx^{(i)})$ from the true $p_{\boldsymbol{\theta}}(\rvz,\rc|\rvx^{(i)})$, and the second term $\mathcal{L}(\boldsymbol{\theta},\boldsymbol{\phi};\rvx^{(i)})$ is 
\begin{align}
    \label{eq:elbo}
       E_{q_{\phi}(\rvz,\rc|\rvx^{(i)})}\left[\log p_{\theta}(\rvx^{(i)},\rvz,\rc)-\log q_{\phi}(\rvz,\rc|\rvx^{(i)})\right],
\end{align}
known as the evidence lower bound.
Since the KL-divergence is non-negative and the value of $\log p_{\boldsymbol{\theta}}(\rvx^{(i)})$ does not depend on $\boldsymbol{\phi}$, minimizing the KL-divergence amounts to maximizing the evidence lower bound. 
\subsection{The reparameterization method with Gumbel-softmax}
The reparameterization method follows the following two steps.
In step 1, we reparameterize the random variable $\rvz \sim q_{\boldsymbol{\phi}}(\rvz|\rvx^{(i)})$ with a deterministic and differential transformation $g_{\boldsymbol{\phi}}(\boldsymbol{\epsilon},\rvx^{(i)})$ of a noise variable $\boldsymbol{\epsilon}$:\begin{equation}
    \rvz=g_{\boldsymbol{\phi}}(\boldsymbol{\epsilon},\rvx^{(i)}) \quad \text { where }\quad \boldsymbol{\epsilon} \sim p(\boldsymbol{\epsilon}).
\end{equation}
In step 2,  we estimate the expectation of some function $h(\rvz)$ $w.r.t$  $q_{\boldsymbol{\phi}}(\rvz|\rvx^{(i)})$ by
 \begin{multline}
    \mathbb{E}_{q_{\boldsymbol{\phi}}(\rvz|\rvx^{(i)})}[h(\rvz)]=\mathbb{E}_{p(\epsilon)}\left[h\left(g_{\phi}\left(\boldsymbol{\epsilon}, \rvx^{(i)}\right)\right)\right] \\= \frac{1}{L} \sum_{l=1}^{L} h\left(g_{\phi}\left(\boldsymbol{\epsilon}^{(l)}, \rvx^{(i)}\right)\right)+o_{p}(1)
 \end{multline}
where $\boldsymbol{\epsilon}^{(l)}$ are samples generated from $ p(\boldsymbol{\epsilon})$. In training the recognition model $q_{\boldsymbol{\phi}}$, non-differentiable categorical samples $\vz$ are replaced with Gumbel-Softmax estimators $\vy$. It results to approximating $\nabla _ { \theta } \rvz$ with $\nabla _ { \theta } \rvy$ for the back pass. It has been shown that samples $\vy$ will become one-hot and the Gumbel-Softmax distribution will converge to the Categorical distribution \cite{cVAE}.
$g_{\boldsymbol{\phi}}(\boldsymbol{\epsilon},\rvx^{(i)})$ is given as \begin{align*}
    g_{\boldsymbol{\phi}}(\boldsymbol{\epsilon},\rvx^{(i)})=\frac{\exp \left(\left(\log \left(\mu_{i}\right)+\epsilon_{i}\right) / \tau\right)}{\sum_{j=1}^{k} \exp \left(\left(\log \left(\mu_{j}\right)+\epsilon_{j}\right) / \tau\right)},
\end{align*} 
where $\epsilon_1...\epsilon_k$ are i.i.d samples drawn from Gumbel (0,1) distribution, $\mu_i$ are the probability of belonging to classes $i$ conditioning on $\rvx^{(i)}$ and $\tau$ is the softmax temperature.
When $\tau \to 0$,
\begin{align*}
    p(\boldsymbol{\epsilon}) \prod_i d\epsilon_i =q_{\boldsymbol{\phi}}(\rvy|\rvx) \prod_i d\ry_i
    \to q_{\boldsymbol{\phi}}(\rvz|\rvx) \prod_i d\rz_i
\end{align*}
and
\begin{multline}
    \label{eq:etm}
\int q_{\phi}(\rvy | \rvx) f(\rvy) d \rvy
=\int p(\boldsymbol{\epsilon}) f\left(g_{\phi}(\boldsymbol{\epsilon}, \rvx)\right) d \boldsymbol{\epsilon}\\= \frac{1}{L} \sum_{l=1}^{L} f\left(g_{\phi}\left(\boldsymbol{\epsilon}^{(l)}, \rvx\right)\right) +o_p(1).
\end{multline}
for some function $f(\rvy)$ corresponding with $h(\rvz)$.
\subsection{Clustering with variational models}
\label{sec:dnns}
To perform clustering embedded in training VAEs, we optimize the lower bound $\mathcal{L}(\boldsymbol{\theta},\boldsymbol{\phi};\rvx^{(i)})$ $w.r.t.$ recognition model
parameters $\boldsymbol{\phi}$ and generative parameters $\boldsymbol{\theta}$ and assign clusters simultaneously. The value of the evidence lower bound (\ref{eq:elbo}) can be wrote as,
\begin{multline}
\label{eq:elbbo}
    \mathcal{L}(\boldsymbol{\theta},\boldsymbol{\phi};\rvx^{(i)})=E_{q_{\boldsymbol{\phi}}(\rvz,\rc|\rvx^{(i)})}[\log p_{\boldsymbol{\theta}}(\rvx^{(i)}|\rvz) +\log p_{\boldsymbol{\theta}}(\rvz|\rc)\\ + \log p_{\boldsymbol{\theta}}(\rc) - \log q_{\boldsymbol{\phi}}(\rvz|\rvx^{(i)}) - \log q_{\boldsymbol{\phi}}(\rc|\rvz)].
\end{multline}
With the approximation (\ref{eq:etm}) to (\ref{eq:elbbo}), (\ref{eq:elbo}) is estimated with Stochastic Gradient Variational Bayes (SGVB) estimators and optimized by Auto-Encoding Variational Bayes (AEVB) algorithm \cite{VAE}. The recognition model $q_{\boldsymbol{\phi}}(\vz,c|\rvx^{(i)})$ and generative model $p_{\boldsymbol{\theta}}(\rvx^{(i)}|\rvz)$ are jointly trained by the encoder and the decoder respectively. For each generated sample $\vy^{(i,l)}$ corresponding to each input $\rvx^{(i)}$, we update the classes by
\begin{equation}
    q_{\boldsymbol{\phi}}(\rc|\vy^{(i,l)})=\frac{p_{\boldsymbol{\theta}}(\rc )p_{\boldsymbol{\theta}}(\vy^{(i,l)}|\rc)}{\sum_{c=1}^kp_{\boldsymbol{\theta}}(\rc )p_{\boldsymbol{\theta}}(\vy^{(i,l)}|\rc)}.
\end{equation}
Note that parameters $\boldsymbol{\pi}$ in $p_{\boldsymbol{\theta}}(c)$ and $\boldsymbol{\mu}_z$ in $p_{\boldsymbol{\theta}}(\rvz|\rc)$ are trained as the model parameters. 
Finally, we construct an estimator of the marginal likelihood lower bound of the full $N$ sample data set based on mini-batches M:
\begin{align}
\begin{split}
    \mathcal{L}(\boldsymbol{\theta}, \boldsymbol{\phi} ; \rvx) &= \widetilde{\mathcal{L}}\left(\boldsymbol{\theta}, \boldsymbol{\phi} ; \rvx\right)+o_p(1)\\&=\frac{N}{M} \sum_{i=1}^{M} \widetilde{\mathcal{L}}\left(\boldsymbol{\theta}, \boldsymbol{\phi} ; \rvx^{(i)}\right)+o_p(1)
\end{split}
\end{align}
with the mini-batches $\rvx^M = \{\rvx^{(i)}\}_{i=1}^M$ randomly drawn from the full data set X. It is pointed that the number of samples $L$ for each data point can be set to 1 as long as the mini-batch size M is large enough, e.g. $M$ = 100.
\section{Experiments}
Our work, to the best of our knowledge, is the only one performing deep clustering on binary data representations in the literature. The most related work is the VaDE \cite{Vade}, a deep clustering method that also trains VAEs with a embedded mixture model but focuses on jointly optimize clustering and generation. The performance of our method will be evaluated with classical clustering methods K-means and Gaussian mixture models (GMMs), as well as deep clustering methods on the hand-written digit image dataset MNIST~\cite{mnist}.
\subsection{Evaluation Metric}
It is not trivial to evaluate the performance of clustering algorithm. We follow the same evaluation metric mentioned in \cite{Xie, Vade} to perform a comparison. With a given number of clusters, the clustering accuracy (ACC) is defined as
\[
ACC=\max_{m\in\mathcal{M}}\frac{\sum_{i=1}^NI\{l_i=m(c_i)\}}{N},
\]
where $N$ is the total number of samples, $l_i$ is i-th ground-truth label, $c_i$ is i-th cluster assignment obtained by the model and $\mathcal{M}$ ranges over all possible mappings between predicted labels and true labels. The best mapping can be efficiently computed by the Hungarian algorithm \cite{doi:10.1002/nav.3800020109}. ACC values varies from 0 to 1 and a higher ACC value indicates a more accurate a clustering performance.

To evaluate the compression quality, the peak signal-to-noise ratio (PSNR) is generally used by measuring the distance of the reconstructed image with the original image. The higher the PSNR, the better the quality of the reconstruction. It is noted that acceptable PSNR for wireless transmission is from 20 dB to 25 dB \cite{PSNR}. To see the advantage of our model in low compressed rate scenario, we will evaluate both clustering performance and compression quality in terms of bits per pixel (BPP). Here, BPP is defined by the number of bits of information stored per pixel. More BPP indicates more memory required to store or display the image. 

\subsection{Experiment Settings}
We trained the model on the train set and then test the performance of our best model on the tested set in order to make the performance convincing and applicable in general. In training, we use feedforward artificial neural networks as the encoder and the decoder. All layers are fully connected and followed with a rectified linear unit (ReLU). For optimizer, we use Adam \cite{Adam} to jointly optimize the full set of parameters with $\beta=(0.9, 0.999)$. The learning rate is initialized as $0.001$ and decreases every 10 epochs with a decay rate of 0.9 down to the minimum of $0.0002$.

The true number of classes $K=10$ is assigned as known. We repeated the experiments on the different values of BPP, which represents different compression rates. BPP value varies from the dimension of latent $\rvz$. For example, with $dim(\rvz)=28$, one gray-scale image input will generate the binary code with size $(1, 28)$ after compression, then we will have $28/1024 = 0.02734375$ BPP in this compression step.

Classical clustering method K-means and GMMs are applied directly on raw image pixels with default settings. The results of VaDE will be reported by re-running the code released from the original paper. The result we obtain is somewhat different from the reported one because of different experimental settings and random seeds.

\begin{table*}[ht]
    \centering
    \begin{tabular}{c|cccc}
    \hline
         Method& K-means&GMM&VaDE&VAB \\
         Best Clustering Acuraccy (\%)&55.37&42.22&95.30&71.69 \\
    \hline
    \end{tabular}
    \caption{The clustering performance is compared on the MNIST test data. The performance of VAB is much better than the classical methods, K-means and GMMs. Although it is not comparable with the performance of VaDE, VAB achieves this result at a much lower bits per pixel as shown in Figure 1(b), more suitable for compressed data.}
    \label{tab:my_label}
\end{table*}

\subsection{Experimental Results}
\begin{figure*}[tb]
\centering
\subfloat[Subfigure 1 list of figures text][Clustering performance of VaDE and VAB against BPP]{
\includegraphics[width=0.5\textwidth]{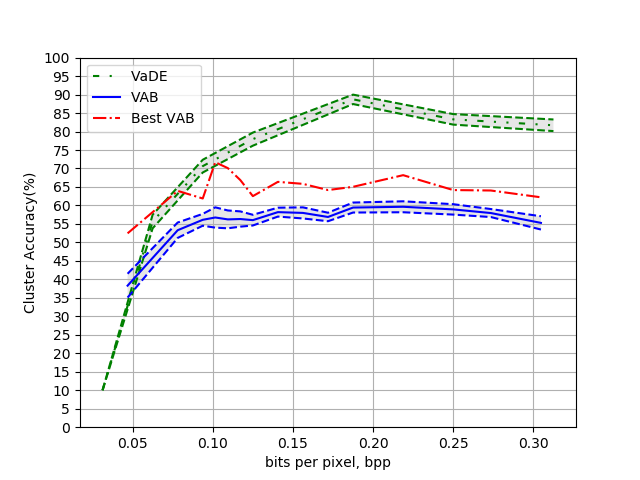}
\label{cluster_acc}}
%\qquad
\subfloat[Subfigure 2 list of figures text][Compression performance of VaDE and VAB against BPP]{
\includegraphics[width=0.5\textwidth]{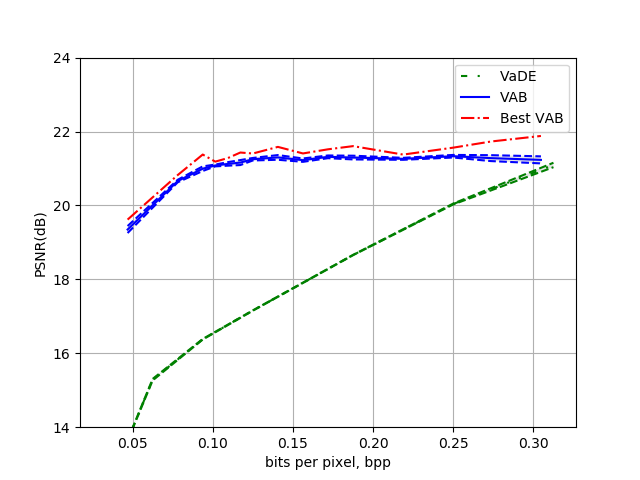}
\label{psnr}}
\caption{The clustering performance and compression performance are shown in (a) and (b) respectively. All results are averaged from 10 experiments and presented by the solid line and the dashed dotted line, showing our method (VAB) and VaDE respectively in both Figures 1(a) and Figure 1(b). The grey area between two dashed lines shows the standard errors of the mean from 10 replications. Figure 1(a) shows that in the low BPP regime, the clustering accuracy of VAB is comparable with VaDE, while its compression performance is much better than VaDE.}
\label{fig}
\end{figure*}
We report all results averaged over 10 experiments corresponding to different BPP values. Figure \ref{fig} shows the clustering performance and the compression quality against BPP respectively. The solid line and the dashed dotted line stands for the mean ACC of our method (VAB) and VaDE respectively in both Figures 1(a) and Figure 1(b), while filling grey area represents the standard errors of the mean over 10 runs. It can be seen that our method can achieve comparable clustering accuracy at a very low bit rate. Meanwhile, the compression quality of our method outperforms the VaDE framework as shown in Figure \ref{fig}(b). 
Table \ref{tab:my_label} presents the clustering accuracy of our method with other baselines. 
Results indicate that our model VAB is more suitable for unsupervised clustering on compressed data, when compared with the state-of-the-art methods.

\section{Discussion}
In this paper, we proposed a new methodology that enables deep clustering in the compressed data domain. The method is presented as a novel amalgamation of Variational Autoencoders (VAEs) with Bernoulli mixture models (BMMs), where VAEs compressed the raw data into representations with generative probability models and BMMs updated clusters of sampled binary representations. Gumbel-Softmax distribution is applied to address the issues caused by discrete values. In the algorithm, we utilized a deep learning architecture to jointly train the model. The optimization target can be treated as a \textit{two-party loss function} consisting of a reconstruction loss and a clustering loss. The learned model greatly improves clustering accuracy compared with well-known clustering methods. Through an approximate mixture of discrete probability models, the proposed solution requires less storage complexity and has the potential to reduce transmission bandwidths. A direction of future works is to develop more learning tools and applications based on compressed data, such as high-dimensional binary data.
% To start a new column (but not a new page) and help balance the last-page
% column length use \vfill\pagebreak.
% -------------------------------------------------------------------------
%\vfill
%\pagebreak
% \clearpage 
\section{Acknowledgements}
The authors thank Professor Yuhong Yang and Feng Qian from the University of Minnesota for valuable discussions.
\balance 
\bibliographystyle{IEEEbib}
\bibliography{reference,addition}

\end{document}